# Monitoring survey of pulsating giant stars in the Local Group galaxies: survey description, science goals, target selection

E Saremi[1,2], A Javadi[2], J Th van Loon[3], H Khosroshahi[2], A Abedi[1], J Bamber[3], S A Hashemi[4], F Nikzat[5] and A Molaei Nezhad[2]

[1] Physics Department, University of Birjand, Birjand 97175-615, Iran
[2] School of Astronomy, Institute for Research in Fundamental Sciences (IPM), Tehran, 19395-5531, Iran
[3] Lennard-Jones Laboratories, Keele University, ST5 5BG, UK
[4] Physics Department, Sharif University of Tecnology, Tehran 1458889694, Iran
[5] Instituto de Astrofisica, Facultad de Fisica, Pontificia Universidad Catolica de Chile, Av. Vicuna Mackenna 4860, 782-0436 Macul, Santiago, Chile

Email: saremi@birjand.ac.ir

**Abstract.** The population of nearby dwarf galaxies in the Local Group constitutes a complete galactic environment, perfect suited for studying the connection between stellar populations and galaxy evolution. In this study, we are conducting an optical monitoring survey of the majority of dwarf galaxies in the Local Group, with the Isaac Newton Telescope (INT), to identify long period variable stars (LPVs). These stars are at the end points of their evolution and therefore their luminosity can be directly translated into their birth masses; this enables us to reconstruct the star formation history. By the end of the monitoring survey, we will have performed observations over ten epochs, spaced approximately three months apart, and identified long-period, dust-producing AGB stars; five epochs of data have been obtained already. LPVs are also the main source of dust; in combination with Spitzer Space Telescope images at mid-IR wavelengths we will quantify the mass loss, and provide a detailed map of the mass feedback into the interstellar medium. We will also use the amplitudes in different optical passbands to determine the radius variations of the stars, and relate this to their mass loss.

## 1. Introduction

Dwarf galaxies are the most abundant type of galaxies in the Universe and so the study of them is of great importance. In this regard, the Local Group provides an excellent opportunity for understanding dwarf galaxies because of their proximity, variety, and wide range of metallicity ($0.002\ Z_\odot$ to $0.08\ Z_\odot$) [1]. The Local Group dwarfs offer the chance for a complete inventory within a galactic environment. We set out to reconstruct their formation histories, and to probe their structure and evolution.

Asymptotic Giant Branch (AGB) stars have a strong influence on the global properties of a galaxy. DUSTiNGS (DUST in Nearby Galaxies with Spitzer) finds that AGB stars are efficient dust producers even down to $0.6\%\ Z_\odot$ [1]. Therefore, they are drivers of galaxy chemical enrichment and evolution via the return of significant amounts of gas and dust to the interstellar medium (ISM) [2]. Several efforts have been made in the last decade to characterize the AGB populations in nearby galaxies [1,3,4]. But only a small fraction of those stars were identified in optical surveys.

Furthermore, these stars are near the end-points of their evolution, and their luminosities directly reflect their birth masses – which is what makes AGB stars powerful probes of a galaxy's star formation history. They trace stellar populations as young as ~ 30 Myr to as old as the oldest globular clusters. The most evolved AGB stars are long period variables (LPVs), and identifying them is one of the best ways to reconstruct the star formation history [5]. LPVs reach the largest amplitudes of their brightness variations at optical wavelengths, due to the changing temperature. We will identify LPVs with an optical monitoring survey of the majority of the Local Group dwarf galaxies, and then will use their luminosity distribution to reconstruct their star formation history employing a method that we have developed and successfully applied in other Local Group galaxies [4,6,7].

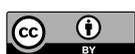

**2. Survey description**
Observations are being made with the Wide Field Camera (WFC) – an optical mosaic camera at the prime focus of the 2.5m Isaac Newton Telescope (INT) in La Palma. The field of view of WFC is about 34'×34' and it covers each galaxy in just one pointing, but dithering between repeated exposures is required to fill the gaps between the detectors. Still, even among the Andromeda system of satellites, none are near enough to one another to fit within one and the same field of view.

LPVs vary on timescales from ~100 days, for low-mass AGB stars to ~1300 days, for the dustiest massive AGB stars. Although we are not aiming to determine accurate periods, but to identify the LPVs and to determine their amplitude and mean brightness, we require monitoring over ten epochs, spaced approximately three months apart. Five epochs of data have been obtained already, and the majority of the targets have been observed more than twice.

We have selected the I-band for the monitoring survey because the spectral energy distribution (SED) of cool evolved stars peak around 1 μm, so they stand out in the I-band and the I-band is where the contrast between the LPVs and other stars is greatest. Also, the bolometric correction – needed to determine the luminosity – in this band is the smallest and the effects of attenuation by dust are minimal. However, we also observe in the V-band, as often as possible, to monitor the variations in temperature – and thus radius – and to further help constrain the SED modelling.

We have chosen exposure times that yield sufficient signal-to-noise (S/N) to detect small changes in magnitude at different epochs. The I-band amplitudes of pulsating AGB stars are > 0.1 mag. Therefore we aim for S/N = 10 for the faintest stars, equivalent to the RGB-tip.

**3. Target selection**
The main targets in this survey are the majority of dwarf galaxies in the Local Group that are visible in the Northern hemisphere, except some that have already been studied, such as NGC147 and NGC185 [8]. These targets contain 55 dwarf galaxies of which 22 are satellites of the Andromeda galaxy – others are satellites of the Milky Way or isolated dwarfs. Also, we have included four distant globular clusters (GC) to investigate the possibility that they are stripped nucleated dwarf galaxies.

We prioritise the targets, principally on the basis of their estimated number of AGB stars; some populous galaxies include IC10 (> $10^4$ AGB stars) and Sextans A and B (> $10^3$ AGB stars). Also, we are attempting to monitor the entire Andromeda system of satellites because these are all accessible to a Northern hemisphere survey. This survey benefits from the homogeneity in distances, completeness, and accuracy. The foreground populations and extinction are modest and similar between all Andromeda satellites. The main aim of this selection is to find out whether the Andromeda system could be a universal template for galaxy evolution, or just one particular case. Individual Milky Way satellites are observed as a comparison to the Andromeda system. Our targets are shown in table 1.

**4. Data**
*4.1. Data reduction*
We have used the THELI pipeline, which is an image processing pipeline with the ability to reduce multi-pointing optical images taken by a mosaic CCD camera. This pipeline works on raw images and removes several instrumental effects, implements photometric calibration and astrometric alignment, and constructs a deep co-added mosaic image complemented by a weight map of the image pixels. In this pipeline more emphasis is put on precise astrometry, than precise photometry [9]. Two examples are shown in figure 1.

**Table 1.** Targets [10].

| Dwarf galaxy | R.A (J2000) | Dec (J2000) | $(m-M)_0$ (mag) | v (mag) | type |
|---|---|---|---|---|---|
| Andromeda I | 00 45 39.8 | +38 02 28 | 24.36±0.07 | 12.7±0.1 | dSph |
| Andromeda II | 01 16 29.8 | +33 25 09 | 24.07±0.06 | 11.7±0.2 | dSph |
| Andromeda III | 00 35 33.8 | +36 29 52 | 24.37±0.07 | 14.4±0.3 | dSph |
| Andromeda IX | 00 52 53.0 | +43 11 45 | 23.89±0.31 | 16.3±1.1 | dSph |
| Andromeda V | 01 10 17.1 | +47 37 41 | 24.44±0.08 | 15.3±0.2 | dSph |
| Andromeda VI | 23 51 46.3 | +24 34 57 | 24.47±0.07 | 13.2±0.2 | dSph |
| Andromeda VII | 23 26 31.7 | +50 40 33 | 24.41±0.10 | 11.8±0.3 | dSph |
| Andromeda X | 01 06 33.7 | +44 48 16 | 24.23±0.21 | 16.6±1.0 | dSph |
| Andromeda XI | 00 46 20.0 | +33 48 05 | 24.40±0.50 | 17.5±1.2 | dSph |
| Andromeda XII | 00 47 27.0 | +34 22 29 | 24.70±0.30 | 18.3±1.2 | dSph |
| Andromeda XIII | 00 51 51.0 | +33 00 16 | 24.40±0.40 | 18.1±1.2 | dSph |
| Andromeda XIV | 00 51 35.0 | +29 41 49 | 24.33±0.33 | 15.9±0.5 | dSph |
| Andromeda XIX | 00 19 32.1 | +35 02 37 | 24.57±0.36 | 15.6±0.6 | dSph |
| Andromeda XV | 01 14 18.7 | +38 07 03 | 24.00±0.20 | 14.6±0.3 | dSph |
| Andromeda XVI | 00 59 29.8 | +32 22 36 | 23.60±0.20 | 15.9±0.5 | dSph |
| Andromeda XVII | 00 37 07.0 | +44 19 20 | 24.50±0.10 | 15.8±0.4 | dSph |
| Andromeda XVIII | 00 02 14.5 | +45 05 20 | 25.66±0.13 | 16.0±9.9 | dSph |
| Andromeda XX | 00 07 30.7 | +35 07 56 | 24.35±0.15 | 18.2±0.8 | dSph |
| Andromeda XXI | 23 54 47.7 | +42 28 15 | 24.67±0.13 | 14.8±0.6 | dSph |
| Andromeda XXII | 01 27 40.0 | +28 05 25 | 24.82±0.31 | 18.0±0.8 | dSph |
| NGC 205 (M110) | 00 40 22.1 | +41 41 07 | 24.58±0.07 | 08.1±0.1 | dE |
| NGC 221 (M32) | 00 42 41.8 | +40 51 55 | 24.53±0.21 | 08.1±0.1 | dE |
| Leo IV | 11 32 57.0 | -00 32 00 | 20.94±0.09 | 15.1±0.4 | dSph |
| Leo V | 11 31 09.6 | +02 13 12 | 21.25±0.12 | 16.0±0.4 | dSph |
| Ursa Major I | 10 34 52.8 | +51 55 12 | 19.93±0.10 | 14.4±0.3 | dSph |
| Ursa Major II | 08 51 30.0 | +63 07 48 | 17.50±0.30 | 13.3±0.5 | dSph |
| WILLMAN 1 | 10 49 21.0 | +51 03 00 | 17.90±0.40 | 15.2±0.7 | dSph |
| Coma Berenices | 12 26 59.0 | +23 54 15 | 18.20±0.20 | 14.1±0.5 | dSph |
| Canes Venatici I | 13 28 03.5 | +33 33 21 | 21.69±0.10 | 13.1±0.2 | dSph |
| Canes Venatici II | 12 57 10.0 | +34 19 15 | 21.02±0.06 | 16.1±0.5 | dSph |
| Bootes I | 14 00 06.0 | +14 30 00 | 19.11±0.08 | 12.8±0.2 | dSph |
| Bootes II | 13 58 00.0 | +12 51 00 | 18.10±0.06 | 15.4±0.9 | dSph |
| Bootes III | 13 57 12.0 | +26 48 00 | 18.35±0.10 | 12.6±0.5 | dSph? |
| Draco | 17 20 12.4 | +57 54 55 | 19.40±0.17 | 10.6±0.2 | dSph |
| Sextans | 10 13 02.9 | -01 36 53 | 19.67±0.10 | 10.4±0.5 | dSph |
| Ursa minor | 15 09 08.5 | +67 13 21 | 19.40±0.10 | 10.6±0.5 | dSph |
| Hercules | 16 31 02.0 | +12 47 30 | 20.60±0.20 | 14.0±0.3 | dSph |
| Leo I | 10 08 28.1 | +12 18 23 | 22.02±0.13 | 10.0±0.3 | dSph |
| Leo II | 11 13 28.8 | +22 09 06 | 21.84±0.13 | 12.0±0.3 | dSph |
| Segue 1 | 10 07 04.0 | +16 04 55 | 16.80±0.20 | 15.3±0.8 | dSph |
| Segue 2 | 02 19 16.0 | +20 10 31 | 17.70±0.10 | 15.2±0.3 | dSph |
| Pisces I | 01 03 55.0 | +21 53 06 | 24.43±0.07 | 14.3±0.1 | dIrr/dSph |
| Pisces II | 22 58 31.0 | +05 57 09 | 21.31±0.18 | 16.3±0.5 | dSph |
| Leo T | 09 34 53.4 | +17 03 05 | 23.10±0.10 | 15.1±0.5 | dIrr/dSph |
| IC 10 | 00 20 17.3 | +59 18 14 | 24.27±0.18 | 09.5±0.2 | dIrr |
| Sagittarius | 19 29 59.6 | -17 40 51 | 25.14±0.18 | 13.6±0.2 | dIrr |
| Cetus | 00 26 11.0 | -11 02 40 | 24.39±0.07 | 13.2±0.2 | dSph |
| WLM | 00 01 58.2 | -15 27 39 | 24.95±0.03 | 10.6±0.1 | dIrr |
| Aquarius | 20 46 51.8 | -12 50 53 | 25.15±0.08 | 14.5±0.1 | dIrr/dSph |
| Leo P | 10 21 45.1 | +18 05 17 | 25.72±0.46 | --- | dIrr |
| Leo A | 09 59 26.5 | +30 44 47 | 24.51±0.12 | 12.4±0.2 | dIrr |
| UGC 4879 | 09 16 02.2 | +52 50 24 | 25.67±0.04 | 13.2±0.2 | dIrr/dSph |
| Pegasus | 23 28 36.3 | +14 44 35 | 24.82±0.07 | 12.6±0.2 | dIrr/dSph |
| Sextans A | 10 11 00.8 | -04 41 34 | 25.60±0.03 | 11.5±0.1 | dIrr |
| Sextans B | 10 00 00.1 | +05 19 56 | 25.60±0.03 | 11.3±0.2 | dIrr |
| Segue 3 | 21 21 31.1 | +19 07 02 | 16.10±0.10 | 14.9±0.5 | GC |
| Sextans C (PAL3) | 10 05 31.8 | +00 04 21 | --- | 14.3±0.7 | GC |
| NGC 2419 | 07 38 07.9 | +38 52 48 | 19.75±0.06 | --- | GC |
| PAL 4 | 11 29 15.8 | +28 58 23 | 20.20±0.01 | --- | GC |

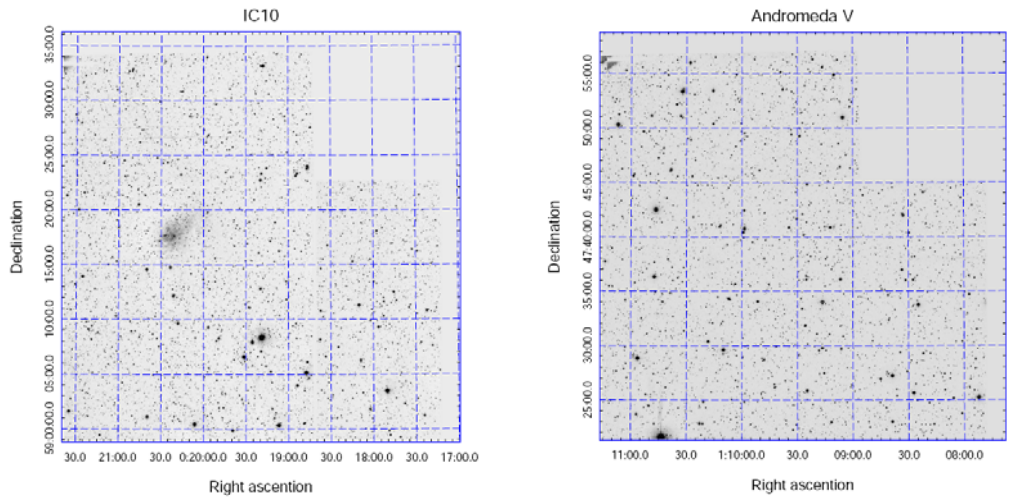

**Figure 1.** The IC10 dwarf galaxy and Andromeda V dwarf galaxy observed with the INT/WFC and processed with THELI.

*4.2. Photometry*
Photometry was obtained for all stars within each frame by automated fitting of a model of the Point Spread Function (PSF), using the DAOPHOT/ALLSTAR software suite [11]. The photometric calibration was performed using standard stars observed on some of the observing nights.

## 5. Science goals
The main objectives of the project are to: identify all LPVs in the dwarf galaxies of the Local Group; obtain accurate time-averaged photometry for all of them; obtain the pulsation amplitude of the LPVs; model their SEDs; study the relation between pulsation amplitude and mass loss (determined from mid-IR observations); reconstruct the star formation history.